\documentclass[conference,a4paper]{IEEEtran}
\IEEEoverridecommandlockouts
\pdfoutput=1
\usepackage{amsmath}
\usepackage{amssymb}
\usepackage{amsfonts}
\usepackage{amssymb,amsmath,amsthm, amsfonts}
\usepackage{epsfig}
\usepackage{subfigure}
\usepackage{calc}
\usepackage{color}
\usepackage[all]{xy}
\usepackage{bm}
\usepackage{cite}
\usepackage{tikz}
\usetikzlibrary{trees}
\usepackage{graphicx}
\usepackage{multirow}

\usepackage{algorithmicx}
\usepackage{algpseudocode}
\usepackage{algorithm}
\usepackage{enumerate}
\usepackage{hyperref}
\usepackage{tikz}
\newcommand*\circled[1]{\tikz[baseline=(char.base)]{
            \node[shape=circle,draw,inner sep=2pt] (char) {#1};}}
\newtheorem{defn}{Definition}
\newtheorem{thm}{Theorem}[section]
\newtheorem{cor}[thm]{Corollary}
\newtheorem{prop}{Proposition}

\newtheorem{lem}[thm]{Lemma}
\newtheorem{conj}[thm]{Conjecture}
\newtheorem{constr}[thm]{Construction}
\newtheorem{note}{Remark}

\newcommand{\bit}{\begin{itemize}}
	\newcommand{\eit}{\end{itemize}}
\newcommand{\bcor}{\begin{cor}}
	\newcommand{\ecor}{\end{cor}}
\newcommand{\beq}{\begin{equation}}
\newcommand{\eeq}{\end{equation}}
\newcommand{\beqn}{\begin{equation*}}
\newcommand{\eeqn}{\end{equation*}}
\newcommand{\bea}{\begin{eqnarray}}
\newcommand{\eea}{\end{eqnarray}}
\newcommand{\bean}{\begin{eqnarray*}}
	\newcommand{\eean}{\end{eqnarray*}}
\newcommand{\ben}{\begin{enumerate}}
	\newcommand{\een}{\end{enumerate}}
\newcommand{\bdefn}{\begin{defn}}
	\newcommand{\edefn}{\end{defn}}
\newcommand{\bnote}{\begin{note}}
	\newcommand{\enote}{\end{note}}
\newcommand{\bprop}{\begin{prop}}
	\newcommand{\eprop}{\end{prop}}
\newcommand{\blem}{\begin{lem}}
	\newcommand{\elem}{\end{lem}}
\newcommand{\bthm}{\begin{thm}}
	\newcommand{\ethm}{\end{thm}}
\newcommand{\bconj}{\begin{conj}}
	\newcommand{\econj}{\end{conj}}
\newcommand{\bconstr}{\begin{constr}}
	\newcommand{\econstr}{\end{constr}}
\newcommand{\bpf}{\begin{proof}}
	\newcommand{\epf}{\end{proof}}

\newcommand{\fQ}{\mbox{$\mathbb{F}_Q$}}
\newcommand{\Zq}{\mbox{$\mathbb{Z}_{2q}$}}
\newcommand{\Zqt}{\mbox{$\mathbb{Z}_{2q}^t$}}
\newcommand{\xyin}{\mbox{$x \in \mathbb{Z}_{2q}, y \in [t]$}}
\newcommand{\xyfail}{\mbox{$(x_1,y_1)$}}

\newcommand{\axyz}{\mbox{$A(x,y; \underline{z})$}}
\newcommand{\axoyoz}{\mbox{$A(x_0,y_0; \underline{z})$}}
\newcommand{\axyzsw}{\mbox{$A(z_y,y; \underline{z}_{(x,y)}  )$}}
\newcommand{\axyzc}{\mbox{$A^c(x,y; \underline{z})$}}
\newcommand{\bxyz}{\mbox{$B(x,y; \underline{z})$}}

\newcommand{\bxyzc}{\mbox{$B^c(x,y; \underline{z})$}}

\newcommand{\cale}{\mbox{${\cal E}$}}

\newcommand{\pz}{\mbox{$P(\underline{z})$}}	
\newcommand{\pxyz}{\mbox{$P_{(x,y)}(\underline{z})$}}

\newcommand{\sigez}{\mbox{$\underline{\sigma}({\cal E}, \underline{z})$}}

\newcommand{\uz}{\mbox{$\underline{z}$}}

\newcommand{\picz}{\mbox{$P({\cal E},\underline{z})$}}

\newcommand{\ez}{\mbox{$({\cal E},\uz)$}}

\newcommand{\txyl}{\mbox{$\theta^{\ell}_{(x,y)}$}}

\newcommand{\calt}{\mbox{${\cal N}$}}
\newcommand{\kstar}{\mbox{$\kappa_{*}$}}

\newcommand{\naa}{\mbox{${\cal N}_{aa} $}}
\newcommand{\nua}{\mbox{${\cal N}_{ua} $}}
\newcommand{\bpc}{\mbox{\text{$B$-plane p-c equations}}}  
\newcommand{\npc}{\mbox{\text{nodal p-c equations}}}  

\newcommand{\bc}{\begin{center}}
	\newcommand{\ec}{\end{center}}

\begin{document}
	
	\title{An Explicit, Coupled-Layer Construction of a High-Rate Regenerating Code with Low Sub-Packetization Level, Small Field Size and $d< (n-1)$}
\author{
  \IEEEauthorblockN{Birenjith Sasidharan, Myna Vajha, and P. Vijay Kumar}
  \IEEEauthorblockA{Department of Electrical Communication Engineering, Indian Institute of Science, Bangalore.\\
    Email: \{birenjith, mynaramana, pvk1729\}@gmail.com}
}	
	\maketitle
	
	%\tableofcontents
	
\begin{abstract}
This paper presents an explicit construction for an $((n=2qt,k=2q(t-1),d=n-(q+1)), (\alpha = q(2q)^{t-1},\beta = \frac{\alpha}{q}))$ regenerating code (RGC) over a field $\mathbb{F}_Q$ having rate $\geq \frac{t-2}{t}$. 
%{\color{red} operating at the Minimum Storage Regeneration (MSR) point.} 
The RGC code can be constructed to have rate $k/n$ as close to $1$ as desired, sub-packetization level $\alpha \leq r^{\frac{n}{r}}$ for $r=(n-k)$, field size $Q$ no larger than $n$ and where all code symbols can be repaired with the same minimum data download.   
%This is the first-known construction of such an MSR code for $d<(n-1)$.  
%More specifically, the parameters of the MSR code can be expressed in terms of two auxiliary parameters $\{q \geq 2,t \geq 2\}$ as follows: $n=qt$, $k=q(t-1)$, $d=(n-1)$, $\alpha=q^t=(r)^{\frac{n}{r}}, \ \beta=q^{t-1}$ and $Q \leq n$, where $R=\frac{t-1}{t}$ is the rate of the MSR code.   
%A building block appearing in the construction is a scalar MDS code of block length $n$.
%and the construction is agnostic to the particular choice of MDS code.  
%The code has a simple layered structure with coupling across layers, that allows both node repair and data recovery to be carried out by making multiple calls to a decoder for the scalar MDS code. 
\end{abstract}

%============INTRODUCTION============================	
\section{Introduction\label{sec:intro}}
	
In an $((n,k,d), (\alpha,\beta))$ regenerating code~\cite{DimGodWuWaiRam} over the finite field  $\mathbb{F}_Q$, a file of size $B$ over \fQ\ is encoded and stored across $n$ nodes in the network with each node storing $\alpha$ coded symbols.  The parameter $\alpha$ is termed as the {\em sub-packetization} level of the code. A data collector can download the data by connecting to any $k$ nodes.  In the event of node failure, node repair is accomplished by having the replacement node connect to any $d$ nodes and downloading $\beta \leq \alpha$ symbols from each node. The quantity $d\beta$ is termed the {\em repair bandwidth}.  The focus here is on exact repair, meaning that at the end of the repair process, the contents of the replacement node are identical to that of the failed node. 

%	\bea \label{eq:cut_set_bd}
%	B & \leq & \sum_{\ell =1}^{k} \min\{\alpha,(d-\ell +1)\beta\} .
%	\eea	
It is well known that the file size $B$ must satisfy the upper bound (see~\cite{DimGodWuWaiRam}): $B \ \leq \ \sum_{\ell =1}^{k} \min\{\alpha,(d-\ell +1)\beta\}$. It follows from this that $B \leq k\alpha $ and equality is possible only if $\alpha \leq (d-k+1)\beta$.   
%==============Literature====================	
\subsection{Literature on MSR Codes}
A regenerating code is said to be a Minimum Storage Regenerating (MSR) code if $B =\alpha k$ and $\alpha =(d-k+1)\beta$, since the amount $n \alpha$ of data stored for given file size $B$ is then the minimum possible. 

The definition of an MSR code requires that all nodes be repairable with the same minimum data download.  There are papers however in the literature that refer to a code as being an MSR code even if the data download is a minimum only for the repair of {\em systematic} nodes.  We will distinguish between the two classes by referring to them as all-node-repair and systematic-repair MSR codes respectively. 

%A construction for all-node-repair MSR codes with $d = n -1 \geq 2k - 1$ is presented in~\cite{SuhRam} that builds on the systematic-repair codes constructed in \cite{ShaRasKumRam_ia}. In \cite{CadHuaLi}, a construction of systematic-repair MSR codes is given, that makes use of permutation matrices. 
	
Several constructions of MSR codes can now be found in the literature.  The product-matrix construction \cite{RasShaKum_pm}, provides MSR codes for any $2k-2 \leq d \leq n-1$.   In~\cite{PapDimCad}, high-rate MSR codes with parameters $(n, k=n-2, d=n-1)$ are constructed using Hadamard designs. In~\cite{TamWanBru}, high-rate systematic-repair MSR codes, known as zigzag codes, are constructed for $d = n-1$. This was subsequently extended to include the repair of parity nodes as well in~\cite{WanTamBru_allerton}.  In \cite{CadJafMalRamSuh}, Cadambe et al. show the existence of high-rate MSR codes for any value of $(n,k,d)$ as $\alpha$ scales to infinity.

Desirable attributes of an MSR code include an explicit construction, high-rate, low values of sub-packetization level $\alpha$ and small field size.  While zigzag codes allow arbitrarily high rates to be achieved, a level of sub-packetization that is exponential in $k$ is required.  In a subsequent paper~\cite{WanTamBru_long}, a systematic-repair MSR code having $\alpha = r^{\frac{k}{r+1}}$ is constructed.  A lower bound $2\log_2 \alpha (\log_{\left(\frac{r}{r-1}\right)} \alpha +1) + 1 \ \geq \ k$ on $\alpha$ is presented in \cite{GopTamCal}. A second lower bound on $\alpha$, $\alpha \geq r^{\frac{k}{r}}$, can be found in \cite{TamWanBru_access_tit}, that applies to a subclass of MSR codes known as help-by-transfer (also known in the literature as access-optimal) MSR codes. For help-by-transfer MSR codes, the number of symbols transmitted as helper data over the network is equal to the number of symbols accessed at the helper nodes.   Prior to this in~\cite{CadHuaLiMeh}, the authors presented a construction of a systematic-repair MSR code that permits rates in the regime $\frac{2}{3} \leq R \leq 1$, and that has an $\alpha$ that is polynomial in $k$. In \cite{RavSilEtz}, explicit help-by-transfer systematic-repair MSR codes are presented with sub-packetization meeting the lower bound $\alpha \geq r^{\frac{k}{r}}$. However the constructions were limited for $r=2, 3$.  In \cite{AgaSasKum}, explicit help-by-transfer systematic-repair MSR codes are presented with sub-packetization meeting the lower bound $\alpha \geq r^{\frac{k}{r}}$ for any $k, r$. In \cite{SasAgaKum}, a high-rate MSR construction for $d=n-1$ is presented that has sub-packetization level $r^{\frac{n}{r}}$ and where all nodes are repaired with minimum data download.  The construction provided was however, not explicit, and required large field size. This is extended for general $k \leq d \leq n-1$ in \cite{RawKoyVis_msr}. In \cite{GopFazVar}, the authors provide a construction for a systematic-repair MSR code for all $k \leq d \leq n-1$, but these constructions are also non-explicit and require large field size. Though suboptimal in terms of repair bandwidth, a vector-MDS code supporting a family of $\alpha = r^p, p \geq 1$ and efficient node-repair is presented in \cite{GurRaw}. 

Most recently, in \cite{YeBar_2}, Ye and Barg present an explicit construction of a high-rate MSR code having rate $k/n$ as close to $1$ as desired, sub-packetization level $\alpha = r^{\frac{n}{r}}$ for $r=(n-k)$, field size $Q$ no larger than $n$, $d=(n-1)$ and where all code symbols can be repaired with the same minimum data download.  Essentially the same construction was rediscovered, albeit some two months later, by the authors of the present paper in \cite{SasVajKum_arxiv}.  The construction in \cite{SasVajKum_arxiv} builds on the earlier construction in \cite{SasAgaKum}. The authors of \cite{GurRaw} observe that the construction in \cite{YeBar_2} can be extended for $d < n-1$ using the technique suggested in \cite{RawKoyVis_msr}, resulting in a non-explicit construction. In \cite{YeBar_1}, the authors present explicit MSR code constructions for $d<n-1$ that requires sub-packetization level $(d-k+1)^{n-1}$.

%
%The combinatorial structure of the parity-check matrix in both \cite{YeBar_2} and \cite{SasVajKum_arxiv} is introduced in the high-rate non-explicit MSR code construction provided in \cite{SasAgaKum}. 
\subsection{Our Contribution} 
In the present paper, we show how the Coupled-Layer MSR code construction in \cite{YeBar_2} (or \cite{SasVajKum_arxiv}) can be modified to handle the case when $d<(n-1)$ to yield an RGC\footnote{In an earlier version of this paper\cite{SasVajKum2}, presented at ISIT 2017, it was incorrectly claimed that the constructed RGC was an MSR code.  However, the construction yields a code whose file size $B < \alpha k$ and thus does not meet the requirements of being an MSR code.} having parameters: 
\bean
(n=2qt,k=2q(t-1),d=n-(q+1)), \\ 
(\alpha = q(2q)^{t-1},\beta = \frac{\alpha}{q}), 
\eean
over a field $\mathbb{F}_Q$ having rate $\geq \frac{t-2}{t}$. A smaller value of $d$ is appealing in practice because it provides greater flexibility in handling node repair.  For instance, it allows one to avoid calling upon nodes that are either slow to respond or else, are otherwise occupied.    
%We adopt the notation introduced in \cite{SasVajKum_arxiv}. 

%=============DESCRIPTION OF THE CODE =========================	
\section{Description of the RGC}
	
\subsection{Code Parameters} 
	
Let $q\geq 2, t \geq 2$ be integers.  Let ${\mathbb{Z}_{2q}}$ denote the set of integers modulo $2q$, $[t]$ denote the set $\{1,2,\cdots,t\}$ and $[0, 2q-1]$ denote the set of integers $\{0, 1, \cdots, 2q-1 \}$.  We describe below the construction of an $\{(n,k,d), (\alpha,\beta)\}$ high-rate RGC over a finite field $\mathbb{F}_Q$ having parameters 
\bean
\left( \ n=2qt, \ k=2q(t-1), \  d=n-(q+1) \ \right),  \\ 
\left(\alpha= q\cdot(2q)^{t-1} , \beta=(2q)^{t-1} \right) \text{  and  } \ \ Q \leq n \ .
\eean 
The file size $B$ of the RGC is such that the rate $R:=\frac{B}{n\alpha}$ of the RGC satisfies:
\bean
 R & \geq & \frac{t-2}{t}. 
\eean
The code is not however an MSR code as it does not meet the requirement $B=k \alpha$. We note that through shortening, we can obtain RGCs having $(n,k,d)=(n-\Delta_s,k-\Delta_s,d-\Delta_s)$ for $0 \leq \Delta_s \ \leq 
k-1$.   Through puncturing, we can obtain RGCs having $(n,k,d)=(n-\Delta_p,k,d)$ for $0 \leq \Delta_p \leq n-d-1$.  A few example parameters are given in the table below:   

\begin{center}
\begin{tabular}{||c||c|c|c|c||}  \hline 
	$(q,t), \Delta_s/\Delta_p$ & \multicolumn{4}{c||}{Parameter set} \\ \cline{2-5}  
	& $n$ & $k$ & $d$ & $\alpha$  \\ \hline 
	$(2,3)$ & 12 & 8 & 9 & 32 \\
	\hline
	$(2,3), \Delta_p=1$ & 11 & 8 & 9 & 32 \\
	\hline
	$(2,3), \Delta_s=2$ & 10 & 6 & 7 & 32 \\
	\hline
	$(2,4)$ & 16 & 12 & 13 & 128 \\
	\hline
	$(3,4)$ & 24 & 16 & 20 & 648 \\
	\hline
\end{tabular} 
\end{center}

%{\color{blue} Given a vector \uz, it will at times be found convenient to have separate access to the $y$th component, $z_y$, $y \in[t]$ of \uz.  For the reason, we define $\pi_y(\uz)  =  \zyz$.   We employ the notation $\pi(\cdot)$ since this is a permutation of the components of \uz.   We will write either \zyz\ or \pizyz\ depending upon whether or not we wish to draw attention to the particular component $z_y$. }
	
	%============Data Cube ===================
\subsection{The Data Cube}
	
\begin{figure}[h!]
			\begin{center}
				\subfigure[The data cube containing $( (2q \times t) \times (2q)^t ) $ symbols over the finite field \fQ. In this example, $2q=4, t=5$. ]{\label{fig:cube}\includegraphics[width=1.5in]{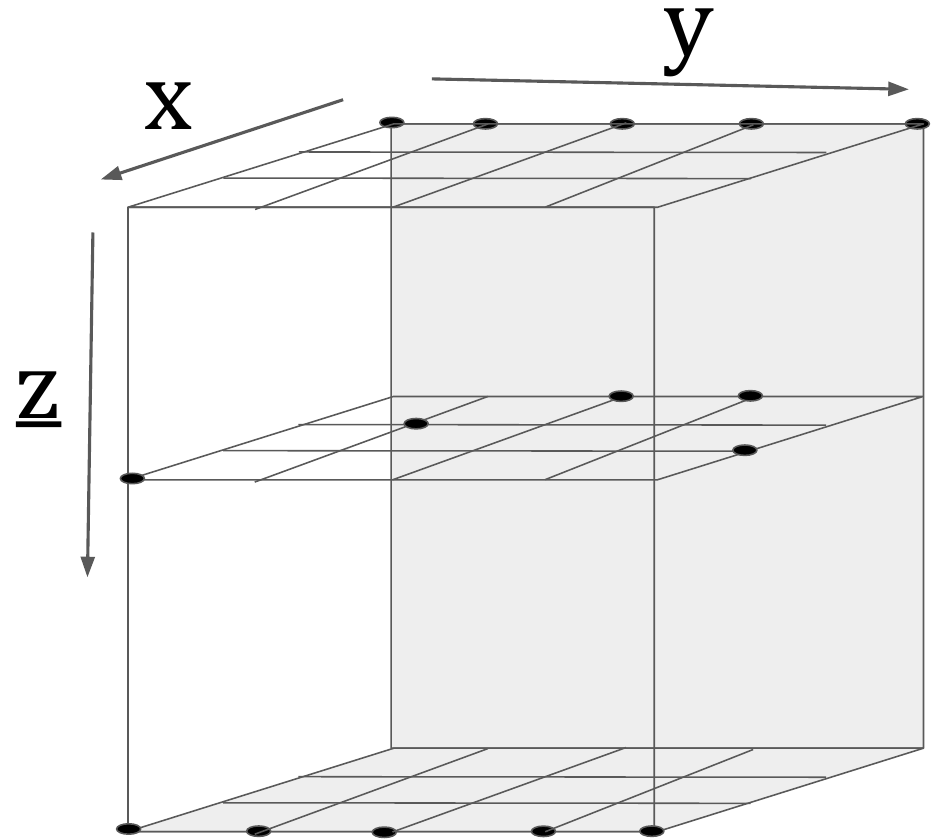}}
				\hspace{0.05in}
				\subfigure[We employ a dot notation to identify a plane. The example indicates the plane $\uz\ = (3,2,0,0,0)$.]{\label{fig:dot_rep}\includegraphics[width=1.5in]{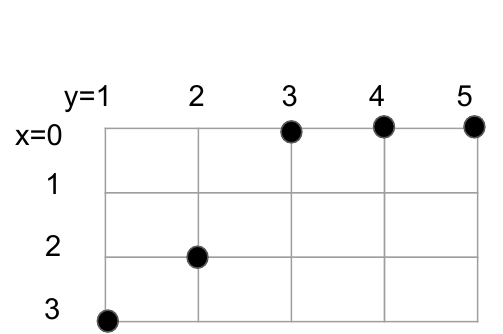}}
				\caption{Illustration of the data cube.\label{fig:repair}}
			\end{center}
		\end{figure}

	The RGC constructed here can be described in terms of an array of symbols over \fQ\ as given below:
	\bean
	{\cal A} & = & \left\{ \axyz \mid x \in \mathbb{Z}_{2q}, y \in [t], \underline{z} \in \mathbb{Z}_{2q}^t \right\} .
	\eean
	This array can be depicted as a  {\em data cube}, see Fig.~\ref{fig:cube} of size $(2q \times t \times (2q)^t)$.
In the figure, the cube appears as a collection of $(2q)^t$ planes, with each horizontal plane indexed by the parameter \uz.  

	From the point of view of the RGC, the data cube corresponds to the data contained in a total of $n=2qt$ nodes, where each node is indexed by the pair of variables:
	\bean
	\left\{ (x,y) \mid x \in \mathbb{Z}_{2q}, y \in [t] \ \right\}  .
	\eean
	The $(x,y)$th node stores the $\alpha_0=(2q)^t$ symbols 
	\bea
	C(x,y) & = & \left\{ \axyz \mid \underline{z} \in \mathbb{Z}_{2q}^t \right\} .  \label{eq:code_and_3D_array}
	\eea  
	Thus each codeword in the RGC is made up of the $n=2qt$ vector code symbols 
    $(C(x,y) \mid \xyin )$, in which each vector has $(2q)^t$ components indexed by $\uz$. It will be explained in Sec.~\ref{sec:alpha} how the $\alpha_0$ components in a vector are mapped to $\alpha$ symbols of a node in the RGC. Let $\Theta$ be a Vandermonde matrix that forms a parity-check matrix of an $[n,k]$-MDS code ${\cal J}$ .  This can be constructed using field size $n$.
    %For example, $\Theta$ could be a Vandermonde matrix, or have form $[P \mid I]$ for $P$ a Cauchy matrix and $I$ an identity matrix, both of which can be constructed using field size $n$.   
 %   Let the rows and columns of $\Theta$ be indexed by $\ell \in [0, q-1]$ and  respectively. 
    We denote by \txyl the  entry of $\Theta$ at the location $(\ell, (x,y))$,  $\ell \in [0, 2q-1]$, $(x,y) \in \mathbb{Z}_{2q} \times [t]$ . Let $u \in \fQ$ satisfy $u \neq 0, u^2 \neq 1$.
    
	By a slight abuse of notation, we will refer to the symbols $\axyz$ as code symbols (as opposed to calling them components of a code symbol) as most of our discussion will involve the symbols $\axyz$. 
	%==================Parity-Check Equations=========================	
	\subsection{Companion Terms, Transformed Code Symbols}
Let us define
\bean
\resizebox{1.0\hsize}{!}{$
\underline{z}_{(x,y)} \ = \  \left\{ \begin{array}{rl}  (x,z_{2},\cdots,z_t), & y=1, \\ 
(z_1,\cdots,z_{y-1},x,z_{y+1},\cdots,z_t), &  2 \leq y \leq t-1, \\
(z_1,z_{2},\cdots,z_{t-1},x), & y=t,\\
\end{array} \right. $}
\eean
in other words, $\underline{z}_{(x,y)} $, is obtained by replacing the $y$th component of \uz\ by $x$.  We next, set 
\bean
\axyzc & = & \axyzsw ,
\eean
and regard $\{\axyz, \axyzc   \}$ as a set of paired elements and \axyzc\ as the {\em companion} of \axyz.  Conversely, \axyz\ is the companion of \axyzc.   Note however, that if $z_y=x$, then $\axyzc = \axyz$ and the element \axyz\ is paired with itself.  For \uz\ such that $z_y \neq x$, we introduce the {\em transformed code symbols} \ \bxyz, \bxyzc:
\bean
\left[  \begin{array}{c} \bxyz \\ \bxyzc \end{array} \right]  & = & 
\left[ \begin{array}{cc} 1 & u \\ u & 1  \end{array} \right] 
\left[  \begin{array}{c} \axyz \\ \axyzc \end{array} \right] ,
\eean
where the inverse transformation is given by 
\bean
\left[  \begin{array}{c} \axyz \\ \axyzc \end{array} \right]  & = & 
\frac{1}{1-u^2}\left[ \begin{array}{cc} 1 & -u \\ -u & 1  \end{array} \right] 
\left[  \begin{array}{c} \bxyz \\ \bxyzc \end{array} \right] .
\eean
If however, $z_y=x$, we simply define
\bean
\bxyz \ = \ \bxyzc \ = \ \axyz \ = \axyzc .
\eean
It can be verified that all $4$ elements $\{\bxyz, \bxyzc, \axyz, \axyzc\}$ can be determined from any $2$ of them.  
 	\begin{figure}[h!]
 		\vspace{-10pt}
		\begin{center}
				\includegraphics[width=1.1in]{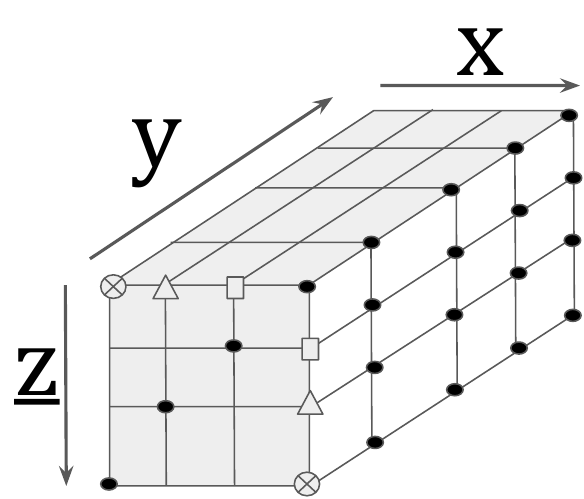}
				\caption{Illustrating $3$ sets of paired symbols $(\axyz, \axyzc)$.} 
				\label{fig:symbol_coupling}
		\end{center}
		\vspace{-10pt}	
	\end{figure}

\subsection{Parity-Check Equations} 
The parity-check (p-c) equations required to be satisfied by the symbols \axyz\ are of two types: {\em $B$-plane p-c equations} and {\em nodal p-c equations}. 

The $B$-plane p-c equations are expressed in terms of the transformed code symbols \bxyz\ and are given by:	\bea
\resizebox{0.9\hsize}{!}{$
	\sum\limits_{x \in \mathbb{Z}_{2q}} \sum\limits_{y \in [t]}   \txyl \bxyz \ = \   0, \ \underline{z} \in \mathbb{Z}_{2q}^t, \ \ell \in [0,2q-1].  $}\label{eq:bpp}
	\eea
Thus there are in all, $(2q)\times (2q)^t$ \bpc\  with $2q$ equations indexed by the parameter $\ell$ per plane \uz. 

The \npc\ involve only the symbols \axoyoz\ lying within the same node. For fixed $(x_0,y_0) \in \mathbb{Z}_{2q} \times [t]$, there are a total of $\left( q\times (2q)^{t-1} \right)$ equations of the form
\bea
\resizebox{0.9\hsize}{!}{$
\axoyoz \theta_{(x_0,y_0)}^{\ell} \ + \ u \sum\limits_{\substack{z'_{y_0} \neq x_0,\\ z'_i = z_i, i \neq y_0} }A(x_0,y_0;\uz') \theta_{(z'_{y_0},y_0)}^{\ell} \ = \  0,$} \label{eq:nodal_1}
\eea
obtained by varying $\ell$, over $0 \leq \ell \leq (q-1)$ and varying $z_i, 1 \leq i \leq t, i \neq y_0$ over all of $\mathbb{Z}_{2q}$, with $z_{y_0} = x_0$ fixed.
These can be alternately be described in terms of their companions as given below: 
\bea
\resizebox{0.9\hsize}{!}{$
\axoyoz \theta_{(x_0,y_0)}^{\ell} \ + \ u \sum_{x \neq x_0}A^c(x,y_0;\uz) \theta_{(x,y_0)}^{\ell} \ = \  0,$} \label{eq:nodal_2}
\eea
where the $\left( q\times (2q)^{t-1} \right)$ equations are obtained this time, by varying $\ell$, over $0 \leq \ell \leq (q-1)$ and varying $\uz \in \mathbb{Z}_{2q}^t$ while maintaining $z_{y_0}=x_0$. 

%===============MSR Code Parameters=================================

	\section{Parameters of the Proposed RGC} 
In the sections to follow, it will be shown that the code constructed above, yields an RGC having parameters 
\bean
\resizebox{\hsize}{!}{$
(n=2qt, \ k=2q(t-1), d=n-q-1), \  (\alpha=(2q)^t/2, \beta=(2q)^{t-1}) .
$}
\eean
and having rate $\geq \frac{t-2}{t}$. 

\subsection{The Value of $\alpha$\label{sec:alpha}} 

With respect to the data cube $\{A(x,y;\uz) \mid \xyin, \uz \in \Zq^t \}$, each pair $(x,y)$ identifies a distinct node.  At the outset each node appears to contain $(2q)^t$ symbols leading to $\alpha=(2q)^t$.  However, these symbols are not linearly independent, since they are subject to the nodal parity-check equations \eqref{eq:nodal_1}.  For a given node $(x_0,y_0)$, there are a total of $(2q)^t/2$ parity-check equations corresponding to a parity-check matrix $J$ having a block-diagonal form: 
\bean
 \underbrace{J_0}_{ (\ (2q)^{t}/2 \ \times \ (2q)^t \ ) }& = & \left[
\begin{array}{cccc}
 \underbrace{J}_{(q \times 2q)} &   &  &  \\
  &  J_0  &  &  \\
  &   &   \ddots & \\
    &   &   & J_0 
\end{array}
\right]
\eean
Each of the matrices $J_0$ is a Vandermonde matrix, hence $J$ has full rank, which means that each node contains just $(2q)^t/2$ linearly independent symbols.   We can thus set $\alpha=(2q)^t/2$. 

\subsection{File Size and Rate of the RGC} 

The total number of parity-check equations, including both $B$-plane p-c equations and nodal p-c equations,  is given by: 
\bean
\underbrace{2qt(2q)^{t-1}q}_{\text{nodal}} + \underbrace{(2q)^t2q}_{\text{planar}} \ = \ (2q)^t(qt+2q) .
\eean
As $\alpha_0$ denotes the number of symbols per node without considering linear dependence among them, we have 
\bean
n \alpha_0 & = & (2qt)(2q)^t.
\eean
It follows that the file size $B$ satisfies the lower bound:
\bean
B & \geq & n \alpha_0 \ - \ (2q)^t(qt+2q) \\
& = & (2qt)(2q)^t  \ - \ (2q)^t(qt+2q) \\
& = & (2q)^t \left\{ q(t-2)\right\} . 
\eean
This leads to the rate bound 
\bean
R & \geq & \frac{t-2}{t}.
\eean
We note that an MSR code having the same parameters would have rate $\frac{k}{n} \ = \ \frac{t-1}{t}$. 

%===============DATA COLLECTION=================================

\section{Pictorial Representation for Planes that Identifies Erased Nodes} 

We associate with each plane \uz, a $(2q \times t)$ $\{0,1\}$ incidence matrix $P(\underline{z})$ given by
\bean
\pxyz  & = & \left\{ \begin{array}{rl} 1 & z_y = x \\
	0 & \text{else}. \end{array} \right.
\eean
Let ${\cal E} = \{  (x_i, y_i) \in \mathbb{Z}_{2q} \times [t] \mid 1 \leq i \leq 2q\}$ denote the location of the $2q$ erased nodes.  Given an erasure pattern \cale\, and a plane \uz\, we define a $(2q \times t)$ $\{0,1\}$ incidence matrix $P(\cale , \underline{z})$ which is the matrix \pz\ with the entries corresponds to the erased nodes circled. For example, if $\cale\ \ = \ \{ (0,2), (1,2), (2,2),(2,4)\}$, with $\uz=[1\ 2 \ 3 \ 1 \ 0]^t$, we obtain: \bean
P({\cal E},\underline{z}) \ =\ \left[ \begin{array}{ccccc} 0 & \circled{0} & 0 & 0 & 1 \\ 1 & \circled{0} & 0 & 1 & 0 \\ 0 &  \circled{1} & 0 & \circled{0} & 0 \\ 0 & 0 & 1 & 0 & 0 \end{array} \right] .
\eean

\subsection{Intersection Score of an Erasure Pattern on a Plane}  Given a plane \uz\ $\in \mathbb{Z}_{2q}^t$ and an erasure pattern ${\cal E}$, we define the {\em intersection score} $\sigma\ez$ to be given by
\bea
\sigma\ez & = & \mid \left\{ y \in [t] \mid (z_y,y) \in \cale \right\} \mid ,  \label{eq:score_def}
\eea
and set $\sigma_{\max}(\cale)=\max \{\sigma(\cale,\uz) \mid \uz \in \mathbb{Z}_{2q}^t\}$.  In terms of the matrix \picz, the intersection score equals the number of circled entries that equal $1$, and hence $\sigma \ez = 1$ in the example above.

\section{Sequential Decoding Approach to Data Collection}  

The data collection property requires that we can recover the data in the presence of $(n-k)=2q$ erasures. Let $\cale= 
\{(x_i,y_i) \mid 1 \leq i \leq 2q\}$ be a fixed erasure pattern. First, we make use of the nodal equations to recover $\alpha$ symbols in each of the $k$ surviving nodes.  Then the aim is to recover the erased code symbols, 
$\{ A(x_i,y_i;\underline{z}) \mid 1 \leq i \leq [t], \uz\ \in \mathbb{Z}_{2q}^t \}$.  We adopt a sequential procedure in which the erased symbols are decoded successively in increasing order of intersection score $s$, $0 \leq s \leq \sigma_{\max}(\cale)$.  The decoding algorithm that relies upon only the $B$-plane p-c equations remains the same as the one described in \cite{SasVajKum_arxiv}. 

%\subsection{Restricting Parity-Check Equations To Erased Symbols} 

\subsection{Case of Zero Intersection Score} Let \uz\ be a fixed plane having intersection score zero. The $2q$ \bpc\ associated to \uz\ are given by
\bean
\sum \limits_{\xyin}   \left\{ A(x,y;\uz) + u A^c(x,y;\uz) \right\} \theta_{(x,y)}^{\ell} & = & 0 . 
\eean 
Since $\sigez=0$, we have that $(z_y,y) \not \in {\cal E}, \text{for any } y \in [t]$.  As a result, the companion symbol \axyzc\ which lies in node $(z_y,y)$, is not erased. It follows that for symbols \axyz\,with $(x,y) \not \in {\cal E}$, both \axyz\ and \axyzc\ are known.   The same argument tells us that for symbols \axyz\,with $(x,y) \in {\cal E}$, while \axyz\ is unknown, \axyzc\ is known. Hence, we can rewrite the parity-check equations associated to plane \uz\ equations in the form $\sum\limits_{(x,y) \in \cale}   A(x,y;\uz) \ \theta_{(x,y)}^{\ell} \ = \ \kappa_{*}$, where $\kappa_{*}$ is generic notion for a known element in the finite field $\mathbb{F}_Q$ that can be determined from the non-erased code symbols.   We are thus left with a set of $2q$ equations involving $2q$ unknowns and a Vandermonde coefficient matrix, so the symbols $A(x,y;\uz)$ lying in a place \uz\ having intersection-score  zero can in this way, be recovered. 

\subsection{Case of Intersection Score $\sigma>0$ }

We show here how one can inductively recover code symbols corresponding to planes \uz\ having intersection score $\sigez>0$, given that symbols in planes $\uz'$ with $\sigma({\cal E}, \uz') < \sigez$ have already been recovered.  
%We have already carried out recovery of code symbols in planes with intersection score $0$, settling the first step of the induction. 

Let an erasure pattern \cale\ and a plane \uz\ be fixed. We first partition the $2q$-erasure location set \cale\ into disjoint subsets, 
\bean
\cale_{0,\underline{z}} & = & \left\{ (x,y) \in \cale \mid x=z_y \right\}, \\
\cale_{1,\underline{z}} & = & \left\{ (x,y) \in \cale \mid (z_y,y) \notin \cale \text{ hence $x \neq z_y$} \right\}, \\
\cale_{2,\underline{z}} & = & \left\{ (x,y) \in \cale \mid (z_y,y) \in \cale, \ x \neq z_y \right\}    .
\eean

It can be verified that in the case of a symbol \axyz\,with $(x,y) \not \in {\cal E}$, the companion symbol \axyzc\ lies either in an unerased node or else in a plane having a lower intersection score, and thus has already been recovered. For this reason, we can assume that the symbols \bxyz\,with $(x,y) \not \in {\cal E}$ are known and the parity-check equations in the inductive decoding process, can once again, be restricted to the erased symbols and their companions, i.e., can be assumed to be of the form
\bean
\sum\limits_{(x,y) \in \cale}   \bxyz\ \ \theta_{(x,y)}^{\ell} & = & \kappa_{*}. 
\eean 
%We first note that the parity-check equations
%\bean
%\sum\limits_{(x,y) \in \cale}  \bxyz \ \theta_{(x,y)}^{\ell} & = & \kappa_{*}. 
%\eean 
These equations allow us to determine the value of the transformed code symbols $\{\bxyz\ \mid (x,y) \in {\cal E} \}$. 

\bit
\item In the case of symbols $\{\bxyz\ \mid (x,y) \in \cale_{0,\underline{z}}\}$, we have $\axyz = \bxyz$ and thus we have recovered the symbols \axyz\ in this instance.  
\item In the case of the symbols $\{\bxyz\ \mid (x,y) \in \cale_{1,\underline{z}}\}$, we have that the complement \axyzc\ does not belong to an erased node and is hence known.  From \bxyz\ and \axyzc\ one can recover \axyz, and so we are done even in this case.  
\item This leaves us only with having to recover symbols $\{\axyz\ \mid (x,y) \in \cale_{2,\underline{z}}\}$.  In the case of such symbols, the companion \axyzc\ can be verified to also belong to a plane having the same intersection score as $\underline{z}$ and hence we can assume that both \bxyz\ and \bxyzc\ have been determined.  From these values, one can determine the value of \axyz.  
\eit 
This concludes the decoding process. 

\section{Node Repair} \label{sec:node_repair}

We turn in this section to node repair and assume node $(x_1,y_1)$ to be the failed node. Since there are a total of $d=n-q-1$ helper nodes, there are a set of $q$ nodes which do not participate in the repair process and which we will term as {\em aloof} nodes. Nodes that are not aloof and which do not correspond to the failed node, will be termed as helper nodes.  
\begin{figure}[h!]
	\begin{center}
		\includegraphics[width=2.2in]{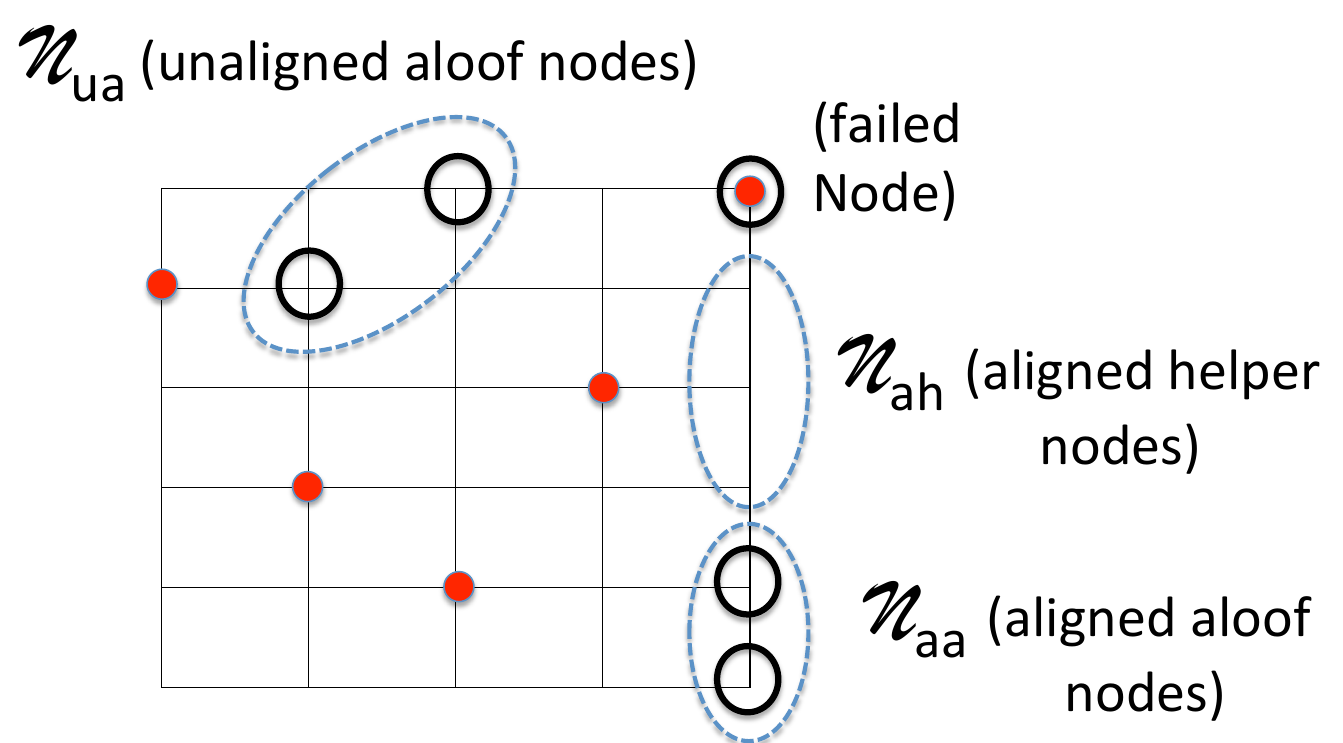}
		\caption{Illustrating the partioning of ${\cal E}$ into aligned (${\cal N}_{aa}$) and unaligned aloof nodes (${\cal N}_{an}$) and aligned helper nodes (${\cal N}_{ah})$.}
		\label{fig:aligned}
	\end{center}	
	%\vspace{-15pt}
\end{figure}
\subsection{Aligned and Unaligned Nodes} We will declare that two nodes to be {\em aligned} if their $y$ coordinates are the same.   Let $\left\{ (x_i,y_i) \ \vert \ 2 \leq i \leq (q+m) \right\}$ denote the coordinates of the helper nodes aligned with $(x_1,y_1)$.  Let us assume that of the $q$ aloof nodes, $(q-m)$ aloof nodes, namely, $\left\{  (x_i,y_i) \ \vert \ q+m+1 \leq i \leq 2q \right\}$, are aligned with the failed node and $m$ of them, namely, $\left\{  (x_i,y_i) \ \vert \ 2q+1 \leq i \leq 2q+m \right\}$, are not aligned.  We set:  
\bean
\resizebox{\hsize}{!}{$
{\cal N}_{ah}  :=  \left\{ (x_i,y_i) \mid i=2, \cdots, (q+m) \right\} \text{(aligned helper nodes)},  
$}  \\
\resizebox{\hsize}{!}{$
{\cal N}_{aa}:=  \left\{  (x_i,y_i) \mid i=q+m+1, \cdots, 2q \right\} \text{(aligned aloof nodes)}, 
$}  \\
\resizebox{\hsize}{!}{$
{\cal N}_{ua}:= \left\{  (x_i,y_i) \mid i=2q+1, \cdots, 2q+m \right\} \text{(unaligned aloof nodes)},
$}  \\
{\cal N} =  (x_1,y_1) \cup {\cal N}_{ah}  \cup {\cal N}_{aa} \cup {\cal N}_{ua}. \hspace*{1in}
\eean
\subsection{The Starting Equations} 

During the repair process, the aloof nodes and the single failed node together behave as though they together constitute a set of $(q+1)$ erased nodes.  For this reason, we set 
\bean
{\cal E} & = & \{ (x_1,y_1)\} \cup \naa \cup \nua ,
\eean
and retain the notation \sigez\ with regard to intersection score. 

While each node $(x,y)$ only stores $\alpha$ non-redundant symbols, it nevertheless has access through computation,  to all $(2q)^t$ symbols $\{\axyz, \uz\ \in \Zqt\}$. Therefore the code does not support help-by-transfer repair. But the only computation required at any helper node is decoding of a half-rate RS code.   During the repair of node \xyfail, we will only call upon the $\beta=(2q)^{t-1}$ symbols $\{ \axyz \mid z_{y_1}=x_1\}$ from a helper node $(x,y)$.  

\subsubsection{Planes with intersection score $1$} Consider first, planes \uz\ which are such that $z_{y_1} =x_1$ and $z_{y_i} \neq x_i$ for any aloof node. Such planes have intersection score $\sigez =1$.  The \bpc\ in such a plane take on the form:
\bea
\sum_{\xyin}  \bxyz \ \txyl & = & 0.  \label{eq:parity_1}
\eea
It can be verified that for $(x,y) \not \in {\cal N}$, the symbols \axyz\ and \axyzc\ are both available for node repair and from these two values, one can compute \bxyz.  Hence we can rewrite \eqref{eq:parity_1} in the form:
\bea 
\sum_{(x,y) \in \calt} \bxyz \ \txyl & = & \kstar.  \label{eq:parity_2}
\eea
For brevity in writing we set:
\bean
a_i \ = \ A(x_i,y_i;\uz), && a^c_i \ = \ A^{c}(x_i,y_i;\uz), \\ 
b_i \ = \ B(x_i,y_i;\uz), && b^c_i \ = \ B^{c}(x_i,y_i;\uz) , \\
\theta_i \ = \ \theta_{(x_i,y_i)}, && \underline{a}^c_{ah} \ = \  [a_2^c, \cdots, a_{q+m}^c]^T, \\
\underline{b}_{aa}  \ = \  [b_{q+m+1}, \cdots, b_{2q}]^T, && \underline{b}_{ua}  \ = \  [b_{2q+1}, \cdots, b_{2q+m}]^T.
\eean
We have the following situation:
\bc
\begin{tabular}{|r|l|} \hline 
Node in ${\cal N}_{ah}$ & $a_i$ known, $a_i^c$ always unknown \\ \hline
Node in ${\cal N}_{aa}$ & $a_i$ unavailable, $a_i^c$ always unknown \\ \hline
Node in ${\cal N}_{ua}$ & $a_i$ unavailable, $a_i^c$ can be unknown \\ \hline
\end{tabular}
\ec
The allows us to rewrite \eqref{eq:parity_2} in the form: 
\bea
\left[ \begin{array}{ccc}  
1 & \cdots & 1 \\ 
\theta_1  & \cdots & \theta_{2q+m} \\
\vdots & \vdots & \vdots \\ 
\theta_1^{2q-1}   & \cdots & \theta_{2q+m}^{2q-1} \\
\end{array} \right] 
\left[ 
\begin{array}{c} 
a_1^c \\
u\underline{a}^c_{ah} \\
\underline{b}_{aa} \\
\underline{b}_{ua} 
\end{array} 
\right]
& = & \kappa_*   .  \label{eq:parity_3}
\eea
Apart from these $2q$ plane-parity equations , we also have the $q$ nodal parity-equations associated to node $(x_1,y_1)$:
\bea
\left[ \begin{array}{ccc}  
1 & \cdots & 1 \\ 
\theta_1  & \cdots & \theta_{2q} \\
\vdots & \vdots & \vdots \\ 
\theta_1^{q-1}   & \cdots & \theta_{2q}^{q-1} \\
\end{array} \right] 
\left[ 
\begin{array}{c} 
a_1^c \\
ua_2^c \\
\vdots \\
ua_{2q}^c \\ \end{array} 
\right]
& = & \kappa_* . \label{eq:nparity_1} 
\eea
Through row-reduction of the parity-check matrix, we can rewrite \eqref{eq:nparity_1} in the form: 
\bea
\resizebox{0.9\hsize}{!}{$
\left[ \begin{array}{c|c|c}  
\underbrace{C_1}_{(m \times q)} & I_m & \underbrace{[0]}_{( m \times (q-m))}  \\ \hline 
\underbrace{C_2}_{(q-m \times q)} & \underbrace{[0]}_{((q-m) \times m)} & I_{q-m} \\
\end{array} \right]   
\left[ 
\begin{array}{c} 
a_1^c \\
ua_2^c \\
\vdots \\
ua_{2q}^c \\ \end{array} 
\right] 
\ = \  \kappa_*  $}. 
\label{eq:nparity_2} 
\eea
%By restricting attention only to those equations involving the unknowns $\{a_1^c,ua_2^c,\cdots, a_{q+m}^c\}$ appearing in \eqref{eq:parity_3}, we obtain: 
%\bea
%\left[ \begin{array}{c|c}  
%\underbrace{C_1}_{(m \times q)} & I_m   \\  
%\end{array} \right]   
%\left[ 
%\begin{array}{c} 
%a_1^c \\
%u \underline{a}^c_{ah} 
%\end{array} 
%\right]
%& = & \kappa_* . \label{eq:nparity_3} 
%\eea
Combining \eqref{eq:parity_3} and first $m$ equations in \eqref{eq:nparity_2} along with further row-reduction, we obtain: 
(see \cite{SasVajKum2_arxiv} for details)
%\bea
%\resizebox{0.9\hsize}{!}{$
%\left[ \begin{array}{c|c|c|c} 
% \multicolumn{4}{c}{\underbrace{V\left( \{ \theta_i^j  \}^{2q-1}_{j=0}, \ i \in [2q+m]  \right) }_{(2q \times (2q+m))} }\\ \hline 
%\underbrace{C_1}_{(m \times q)} & \underbrace{I_m}_{(m \times m)} & \underbrace{[0]}_{( m \times (q-m))} & \underbrace{[0]}_{( m \times m )} 
%\end{array} \right] 
%\left[ 
%\begin{array}{c} 
%a_1^c \\
%u\underline{a}^c_{ah} \\
%\underline{b}_{aa} \\
%\underline{b}_{ua} 
%\end{array} 
%\right]
%\ = \  \kappa_*   $}.  \label{eq:cparity_1}
%\eea
%Further row reduction can be shown to yield: 
\bea
\resizebox{0.9\hsize}{!}{$
\left[ \begin{array}{c|c|c|c} 
 \underbrace{[0]}_{(m \times q)} & \underbrace{[0]}_{(m \times m)} & \underbrace{[0]}_{( m \times (q-m))} &  \underbrace{C_3}_{(m \times m)} \\ \hline 
 \multicolumn{3}{c|}{\underbrace{V\left( \{ \theta_i^j  \}^{2q-1}_{j=0}, \ i \in [2q]  \right) }_{((2q) \times (2q))} } & 
 \underbrace{[0]}_{(2q \times m)} 
\end{array} \right] 
\left[ 
\begin{array}{c} 
a_1^c \\
u\underline{a}^c_{ah} \\
\underline{b}_{aa} \\
\underline{b}_{ua} 
\end{array} 
\right]
\ = \  \kappa_*   $}.  \label{eq:cparity_2}
\eea
Clearly, the matrix on the left is nonsingular since $C_3$ is a Cauchy matrix and it follows therefore that we can recover the unknown vector: 
$[a_1^c, \ u[\underline{a}^c_{ah}]^T, \ [\underline{b}_{aa}]^T, \ [\underline{b}_{ua}]^T]^T$. The vector  $[a_1^c, \ [\underline{a}^c_{ah}]^T]^T$ consists of $(q+m)$ symbols from the same node that participate in the $q$ nodal p-c equations involving $2q$ symbols. Thus we can decode $2q$ symbols $\{ A(x_1,y_1; \underline{z}_{(x,y_1)} \mid x \in \mathbb{Z}_{2q}\}$ belonging to the failed node.
    
The case of planes having intersection score $>1$ can be shown to reduce to the case of plane shaving intersection score $1$ using arguments similar to those employed in describing how data collection is carried out.  For lack of space, we omit the details.

\bibliographystyle{IEEEtran}
\bibliography{isit2017}

\end{document}